\documentclass[dvips]{article}

\usepackage{amssymb,latexsym}
\usepackage{amsmath}
\usepackage{graphicx}
\usepackage{array}
\usepackage{changebar}
\usepackage{vmargin}

\setmarginsrb{1in}{.5in}{1in}{1.in}{14pt}{0.3in}{14pt}{0.494in}

\begin{document}

\title{Simple Model of Sickle Hemogloblin }
\author{Andrey Shiryayev, Xiaofei Li and James D. Gunton
      \\    Department of Physics, Lehigh University, Bethlehem,
      Pa
       18015}
\date{\today}
\maketitle

\begin{abstract}
    \noindent
A microscopic model is proposed for the interactions between
sickle hemoglobin molecules based on information from the
protein data bank.  A Monte Carlo simulation of a simplified
two patch model is carried out, with the goal of understanding
fiber formation.  A gradual transition from monomers to one
dimensional chains is observed as one varies the density of
molecules at fixed temperature, somewhat similar to the
transition from monomers to polymer fibers in sickle
hemoglobin molecules in solution. An observed competition between
chain formation and crystallization for the model is also discussed.
The results of the simulation of the equation of state are shown to be in
excellent agreement with a theory for a model of globular
proteins, for the case of two interacting sites.
\end{abstract}

\maketitle

\section{Introduction}
The condensation of globular proteins from solution is a subject
of considerable experimental and theoretical activity. One one
hand, it is important to grow high quality protein crystals from
solution in order to be able to determine protein structure (and
thus function) from X-ray crystallography.  On the other hand,
many diseases are known to be caused by undesired condensation of
proteins from solution.  In both cases one needs to have a
reasonably detailed model of the protein--protein interactions in
solution in order to predict the phase diagram, condensation rate
and growth kinetics.  This is a major challenge to theorists, as
these protein-protein interactions arise from many sources and are
still relatively unknown in most cases.\\
\\
An important example of undesired protein condensation occurs with
sickle hemoglobin (HbS) molecules in solution.  It is known that
deoxygenated sickle hemoglobin molecules in red blood cells can
undergo a two step nucleation process that leads to the formation
of polymer fibers in the cell \cite{Ferrone_85_01,Ferrone_02_03}.
These fibers distort the cells and make it difficult for them to
pass through the capillary system on their return to the lung. A
direct determination of the homogeneous nucleation of HbS fibers
\emph{in vitro} has shown that the nucleation rates are of the
order of $10^6 - 10^8$ $cm^{-3}s^{-1}$ and that the induction
times agree with Zeldovich's theory \cite{Galkin_04_01}. These
rates are comparable to those leading to erythrocyte sickling
\emph{in vivo}.  They are also approximately nine to ten orders of
magnitude larger than those known for other protein crystal
nucleation rates, such as lysozyme.

\begin{figure}
	\center
    \rotatebox{-90}{\scalebox{.5}{  \includegraphics{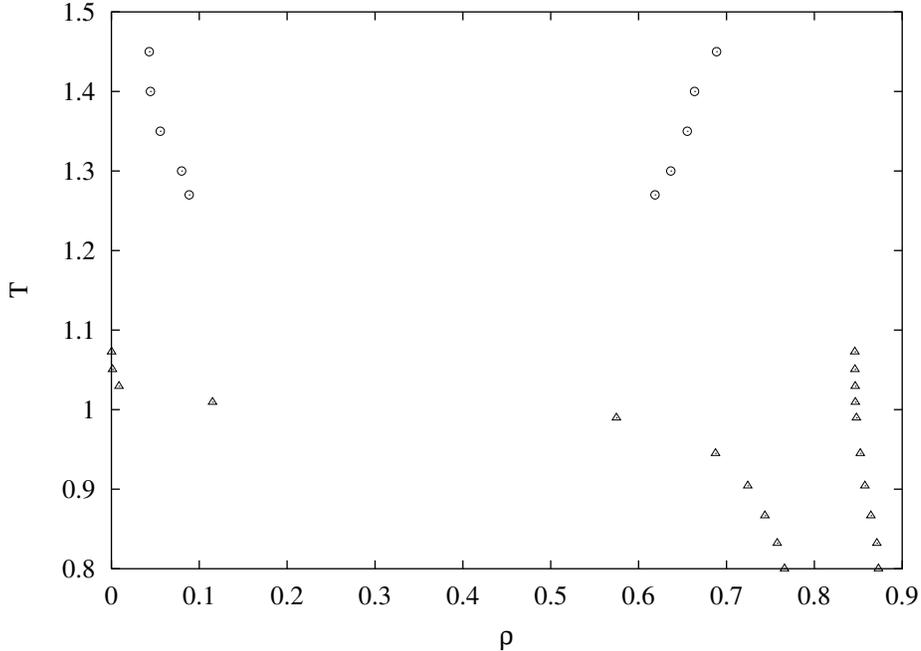}}}
     \caption{Phase diagram of a modified Lennard-Jones model
     including solvent-solute interactions.
      Open triangles denotes liquid-solidus line;
     open circles denotes  fluid-fluid coexistence.
     From \protect\cite{Shiryayev_05_01}. Details of the model are given
     in this reference.}
     \label{fig_solvent}
     \end{figure}

Consequently a goal of current research is to understand at a
molecular level this nucleation process and by controlling the
conditions on which the nucleation depends, to slow down the
nucleation rate such as to prevent the polymerization from
occurring while HbS is in its deoxygenated state in the cells.  To
do this requires understanding the protein-protein interactions,
in order to predict the phase diagram and nucleation rate for
sickle hemoglobin molecules.  The phase diagram for HbS is only
partially known experimentally.  It is known that there is a
solubility line separating monomers and fibers
\cite{Eaton_77_00,Eaton_90_01} and evidence exists for a spinodal
curve with a lower critical point \cite{Palma_91_01}. In a
previous publication \cite{Shiryayev_05_01} we obtained a phase
diagram that was qualitatively similar to this, namely, a
liquid-liquid phase separation with a lower critical point. In
addition, we determined the location of the liquidus and
crystallization lines for the model, as shown in Figure
\ref{fig_solvent}.

However, although yielding a lower solution critical point, this
model was unable to predict the formation of polymer fibers, as it
was based on a spatially isotropic, short range  protein-protein
interaction (e.g. a square well or a modified Lennard-Jones
potential energy).  Fiber formation clearly requires anisotropic
interactions.  In this paper we propose an anisotropic model for
the HbS-HbS interactions, based on an analysis of the contacts for
HbS crystals from the protein data bank. We also define an order
parameter to describe the polymerization of this model.  As the
full model is complex and involves several unknown interaction
parameters, we study a simplified version of the model (a two
patch model) in order to gain some insight into the nature of the
fiber formation. We determine some aspects of the phase diagram
for the two patch model via Monte Carlo simulation and biasing
techniques and show in particular that it yields one dimensional
chains that are somewhat similar to the polymer fibers formed in
HbS nucleation. Real HbS fibers, however, have a diameter of about
21 nm.  In addition, the strands within the fiber are packed into
double strands. Thus the two patch model is too simple to describe
the polymer fiber phase transition observed in HbS.  Future work
will be necessary to obtain reasonable estimates of the
interaction parameters in the full model, in order to obtain a
realistic model for the polymer fiber phase
transition.\\
\\
The outline of the paper is as follows.  In section 2 we propose
an anisotropic interaction model for the pair interactions between
HbS molecules.  In section 3 we define an order parameter that
measures the degree of polymerization in the system.  In section 4
we present the results of our Monte Carlo simulation for a two
patch approximation to the full model, since we are unable to make
realistic estimates for the interaction parameters for the full
model.  A biasing technique is used in order to examine the nature
of the chain formation. In section 5 we summarize the results of a
perturbation theory as applied to an eight patch model and to our
two patch model.  In the latter case we show that the simulation
results are is excellent agreement with this theory. In section 6
we present a brief conclusion and suggest directions for future
research on this subject.

\section{Anisotropic model for the Hemoglobin S polymerization}

\begin{figure}
 \center
     \includegraphics[width=8cm]{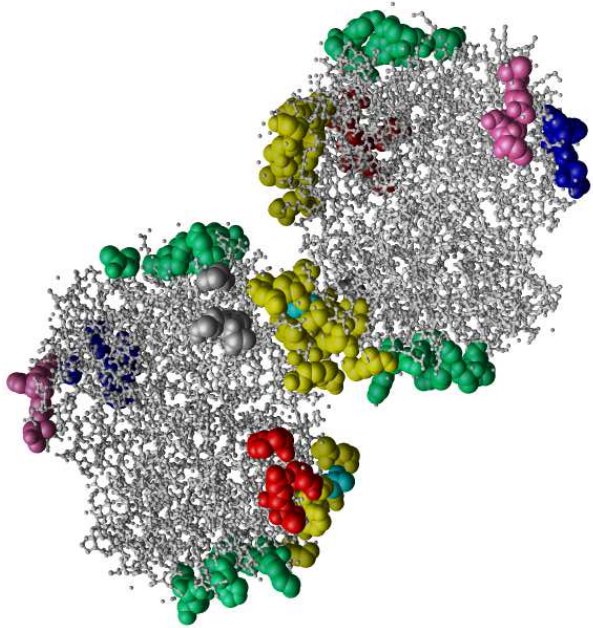}
     \caption{Location of the residues participating in different contacts.
 Yellow areas denote the lateral contacts, green areas denote the axial contacts. }
     \label{fig_HbS_Patches}
     \end{figure}

Protein molecules in general,  and sickle hemoglobin molecules in
particular, are  very complicated objects, typically consisting of
thousands of molecules.
 There are many types of forces between protein molecules in
solution, such as Coulomb forces, van der Waals forces,
hydrophobic interactions, ion dispersion forces and hydrogen
bonding. Although these interactions are complex, considerable
success in predicting the phase diagrams of several globular
proteins in solution has been accomplished by using rather simple
models with spatially isotropic interactions. These models share
in common a hard core repulsion together with a short range
attractive interaction (i.e. the range of attraction is small
compared to the protein diameter). However,in general the
protein-protein interactions are anisotropic,  often arising from
interactions  between specific amino acid residues on the surfaces
of the interacting molecules; i.e., certain areas of a protein
surface interact with certain areas of another molecule's surface.
Thus there has been some recent work attempting to model these
anisotropic interactions. In such models a given protein molecule
is represented by a hard sphere with a set of patches on its
surface
\cite{Wertheim_87_01,Jackson_88_01,Benedek_99_01,Sear_99_01,Curtis_01_01,Kern_03_01}.
Intermolecular attraction is localized on these patches. Typically
the models assume that two protein molecules interact  only when
they are within the range of attractive interaction and when the
vector joining their centers intersects patches on the surface of
both molecules. The fluid-fluid diagram for such a model was
studied by Kern and Frenkel  \cite{Kern_03_01} in a Monte Carlo
simulation and by Sear theoretically \cite{Sear_99_01}. In these
studies all patches are assumed to interact with each other
equally. This approximation gives a good qualitative picture of
the possible importance of anisotropy in protein phase diagrams.
However, when we consider the  behavior of a specific globular
protein,  this approximation is too simple and does not reflect
the actual structure of protein aggregates. This is particularly
important in the fiber formation of HbS molecules in solution.
Figure \ref{fig_HbS_Patches} shows interacting HbS molecules, with
the different pairs of interacting regions in the crystal state of
HbS.  We use this information,  in accordance with the contact
information in the HbS crystal \cite{Padlan_85_01, Adachi_97_00},
to develop a model for the anisotropic interactions between the
HbS molecules.

To develop an anisotropic model to describe these interacting
molecules, we allow for the possibility of different interaction
strengths between different pairs of these interacting patches. To
characterize such interactions, we introduce an interaction matrix
$\{\epsilon_{kl}\}_{mm}$, where $\epsilon_{ij}$ is the strength of
interaction between the $k^{th}$ and $l^{th}$ patches and $m$ is
the total number of patches on a protein surface. This is a
symmetric matrix that describes the strength of interaction
between each pair of patches. In our particular model we choose a
square well attraction between patches, although this is not a
necessary restriction on the interactions.  Thus the interaction
matrix consists of the well depth values for the different
patch-patch interactions. We define the pair potential between two
molecules in this case in a way  similar to
\cite{Sear_99_01,Kern_03_01}, but generalizing to the case of
different patch-patch interactions:
\begin{equation}
U_{i,j}(\textbf{r}_{ij}, \Omega_i, \Omega_j) = U^0_{ij}(r_{ij})
\sum_{k,l}^{m} \epsilon_{kl}
\Theta_k(\hat{\textbf{r}}_{ij}\cdot\hat{\textbf{n}}_{ik})
\Theta_l(-\hat{\textbf{r}}_{ij}\cdot\hat{\textbf{n}}_{jl})
\label{AnisotropicPairInteraction}
\end{equation}
Here $m$ is the number of patches, $\Omega_i$ is the orientation
(three Euler angles, for example) of the $i^{th}$ molecule,
$U^0_{ij}$ is the square well potential for the unit well depth
and $\hat{\textbf{n}}_{ik}$ is the $k^{th}$ patch direction in the
laboratory reference of frame:
\begin{equation}
\hat{\textbf{n}}_{ik} = \tilde{R}(\Omega_i)\hat{\textbf{n}}^0_k
\label{PatchDirection}
\end{equation}
Here $\hat{\textbf{n}}^0_k$ is the $k^{th}$ patch direction in the
$i^{th}$ molecule reference of frame and $\tilde{R}(\Omega_i)$ is
a rotation matrix between the $i^{th}$ molecule and the laboratory
reference of frame. $\Theta(x)$ in
(\ref{AnisotropicPairInteraction}) is a step function
\begin{equation}
\Theta_k(x) = \left\{ \begin{array}{ll} 1 & \mbox{if $x \geq \cos\delta_k$;}\\
0 & \mbox{if $x < \cos\delta_k$.}\end{array} \right.
\label{StepFunc}
\end{equation}
where $\delta_i$ is the half open angle of the $i^{th}$ patch. In
other words, $\Theta(\hat{\textbf{r}}_{ij}\hat{\textbf{n}}_{ik})$
is equal to 1 when the vector joining two molecules intersects the
$k^{th}$ patch on the surface. If patches do not overlap then the
sum in equation (\ref{AnisotropicPairInteraction}) has at most one
term. The radial square well dependence with range $\lambda$ is
given by
\begin{equation}
U^0_{ij}(r) = \left\{\begin{array}{ll}\infty & \mbox{for
$r<\sigma$}\\ -1 & \mbox{for $\sigma \leq r < \lambda\sigma$} \\ 0
& \mbox{for $r \geq \lambda\sigma$} \end{array} \right. \notag
\end{equation}

Until now what we have done is valid for any pair of molecules
interacting through pairwise patch interactions. We now consider
the specific case of the HbS molecule. As noted above, in
agreement with contact information for the HbS crystal
\cite{Padlan_85_01, Adachi_97_00}, this molecule has two lateral
and two axial patches that are involved in intra double strand
contacts and four more patches involved in inter double strand
contacts. Thus we have eight possible patches for the HbS molecule
(Figure \ref{fig_HbS_Patches}). One of the lateral patches
contains a $\beta 6$ valine residue and another lateral patch
contains an acceptor pocket for this residue. But it is known
\cite{Eaton_90_01} that only half the mutated sites are involved
in the contacts in the HbS crystal. Thus we have another possible
set of 8 patches similar to the first one. The total number of
patches is therefore sixteen (two equal sets of eight patches).
The interaction matrix can be built assuming that the first
lateral patch (of any set) can interact only with the second
lateral patch (of any set). The same is true for axial patches.
The remaining four patches in each set can be divided into pairs
in a similar way, in accordance with \cite{Padlan_85_01}. This
gives the following interaction matrix (for one set of eight
patches):
\begin{equation}
\Upsilon = \left( \begin{array}{cccccccccccccccc}
0 & \epsilon_1 & 0 & 0 & 0 & 0 & 0 & 0 \\
\epsilon_1 & 0 & 0 & 0 & 0 & 0 & 0 & 0 \\
0 & 0 & 0 & \epsilon_2 & 0 & 0 & 0 & 0 \\
0 & 0 & \epsilon_2 & 0 & 0 & 0 & 0 & 0 \\
0 & 0 & 0 & 0 & 0 & \epsilon_3 & 0 & 0 \\
0 & 0 & 0 & 0 & \epsilon_3 & 0 & 0 & 0 \\
0 & 0 & 0 & 0 & 0 & 0 & 0 & \epsilon_4 \\
0 & 0 & 0 & 0 & 0 & 0 & \epsilon_4 & 0 \\
\end{array} \right) \label{FullInteractionMatrix}
\end{equation}
where $\epsilon_1$ is the strength of the lateral contact,
$\epsilon_2$ is the strength of the axial contact, $\epsilon_3$
and $\epsilon_4$ are the strength of the inter double strand
contacts. We will refer to this model as the "full model" for HbS.

\section{Polymerization order parameter in a system of patchy hard spheres}

One of the goals of a study of  HbS molecules in solution is to
calculate the free energy barrier that separates the monomer
solution from the aggregate (polymer chains/fibers) state. For
this purpose we have to specify an "order" parameter that measures
the degree of polymerization in the system. The structure of the
aggregate depends strongly on the configuration of patches.
Therefore, to separate the aggregate state from the monomer state,
the order parameter should reflect the configuration of patches.
(Note that the order parameter as defined below is only zero in
the case in which there are only monomers.)  Since in our model
the regions on the molecular surface not covered by patches do not
interact (except through the hard core repulsion), we can measure
the degree of polymerization by measuring the fraction of the
patches involved in actual contacts.

We assume that any two particles at any given time can have no
more than one contact between each other. This condition is a
little stronger than just a non-overlap of the patches. For each
pair of particles we introduce a quantity that shows how much
these particles are involved in polymerization (basically, showing
the presence of the contact between them):
\begin{equation}
\psi_{ij}(\textbf{r}_i, \textbf{r}_j, \Omega_i, \Omega_j) =
\sum_{k,l}^{N_p} w_{kl}
f_k(\hat{\textbf{r}}_{ij}\cdot\hat{\textbf{n}}_{ik})
f_l(-\hat{\textbf{r}}_{ij}\cdot\hat{\textbf{n}}_{jl})
\label{PairOP}
\end{equation}
where $w_{kl}$ is a weight of the contact between the $k^{th}$
patch of the $i^{th}$ molecule and the $l^{th}$ patch of the
$j^{th}$ molecule. We choose the weight matrix to be the
interaction matrix. $f_k(x)$ is equal to $x$ for $x >
\cos\delta_k$ and is zero otherwise. Due to our assumption of only
one contact per pair of particles, the sum in (\ref{PairOP}) has
at most one nonzero term. We next define the order parameter of
one particle to be
\begin{equation}
\psi_i(\textbf{r}_i) = \frac{\sum_{j} \psi_{ij}}{\sum_{k,l}
w_{kl}}
\end{equation}
The term in the denominator is a normalization constant. The order
parameter of the whole system is
\begin{equation}
\psi = \frac{1}{N}\sum_i^N \psi_i \label{PatchyOrderParameter}
\end{equation}

\noindent This choice of order parameter reflects the patch
configuration; the magnitude of the order parameter increases as
the number of contacts in the system increases. It is also
rotationally invariant. However, this construction has its
disadvantages. For some choices of weight matrices it is possible
that a fewer number of contacts could lead to a larger order
parameter if these contacts has significantly larger weights.
However, the choice of the weight matrix equal to the interaction
matrix seems to be natural.

\section{Two-patch model and one-dimensional fiber formation}

The full model described earlier for the sickle hemoglobin
molecule is  complex and has several interaction parameters which
have not yet been determined
 from experiment. Because we are interested in studying the fiber
formation, we use a simplified model that still can produce fiber
chains. We simplify the original model by reducing the number of
contacts. Since one important feature of  HbS fibers is the
presence of twisted, quasi-one dimensional chains,  we consider a
system of particles with only two (axial) patches. This model is
obviously not an accurate description of interacting HbS
molecules, but it can lead to the formation of one dimensional
chains. The interaction matrix for the simplified two-patch model
is just
\begin{equation}
\Upsilon = \left( \begin{array}{cc} 0 & 1 \\ 1 & 0 \end{array}
\right) \label{TwoPatchInteractionMatrix}
\end{equation}

\begin{figure}
\center
\includegraphics[width=8cm]{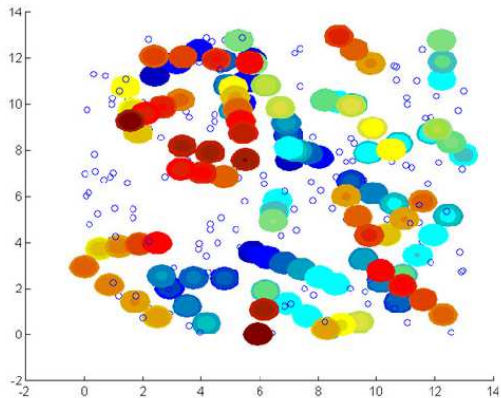}
 \caption{Intermediate state of one dimensional chain formation using the two patch model.
 The actual particle size is equal to the size of the colored spheres.  Particles that
 are not involved in any chains are scaled down and shown as small blue circles.  Particles
 that are part of some chains are shown as colored spheres.  Color shows the depth of the
 particle (z-axis); blue is the deepest and red is the least deep.  Simulations were performed
 at low temperature and high supersaturation.}
 \label{fig_HbS_Chains1}
 \end{figure}

\noindent Since there are two patches on each sphere, there is
only a fluid-solid phase transition.  (The full model described
above can have a fluid-fluid phase transition as well.) The
formation of the one dimensional chains, therefore, is a gradual
transition from the fluid phase as the density of the molecules is
increased. Figure \ref{fig_HbS_Chains1} shows chains that result
in our simulation.

Since the formation of these chains is a gradual transition as one
increases the density, they do not arise
 from homogeneous nucleation. A necessary property of nucleation
is the existence of a nucleation barrier in the free energy
dependence on the order parameter. This barrier should separate
two wells; one well corresponds to the metastable phase, the other
 to the more stable phase.  In our case there will not be such a
barrier.  To examine the nature of the chain formation, we
determined the dependence of the free energy on the patch order
parameter by performing two series of umbrella sampling Monte
Carlo simulations. The first set of simulations was done at
$T=0.185$ and $P=1.0$.
\begin{figure}
\center
\includegraphics[width=8cm]{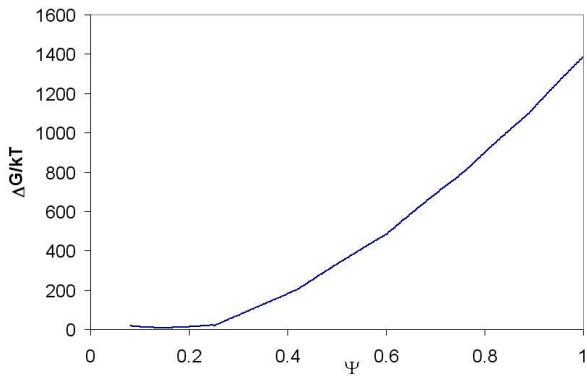}
\caption{Plot of the free energy $\Delta \Omega$ in units of $kT$
versus the order parameter for the patches, $\Psi$, defined in
the text. Simulations are at a temperature $kT=0.185$ and 
pressure $P=1.0$. The  minimum at $\Psi$ around .15 corresponds 
to a liquid state, which is a mixture of monomers and dimers.} 
\label{fig_T_0185_P_1}
\end{figure}

\begin{figure}
\center
\includegraphics[width=8cm]{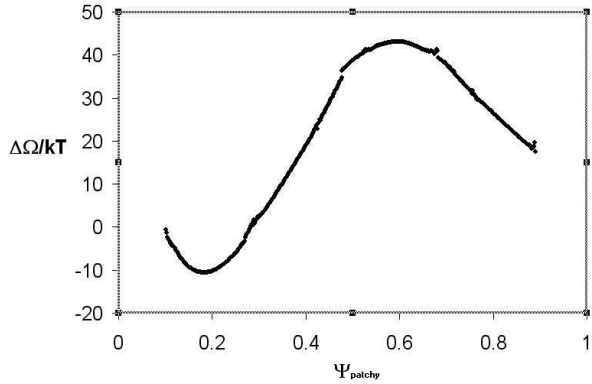}
\caption{Plot of the free energy $\Delta \Omega$ in units of $kT$ versus the order
parameter for the patches, $\Psi$,  defined in the text.  Simulations are
at a temperature $kT=0.185$ and pressure $P=1.6$.  The minimum
at $\Psi$ around 0.2 corresponds to a mix of monomers and dimers, while the minimum
at $\Psi$ close to 1 corresponds to a crystal state in which all the patches 
are in contact.}
\label{fig_TwoPatch_185_160}
\end{figure}

\noindent In this case the initial liquid state does not
crystallize in the absence of the biasing and remains in the
liquid state. The order parameter has only one minimum about
$\Psi_{patch}=\Psi \cong 0.15$, corresponding to a mix of monomers
and dimers (Figure \ref{fig_T_0185_P_1}). As we increase the
pressure to $P=1.6$ the free energy now has a minimum at a lower
value of $\Psi$ that corresponds to the liquid state (figure
\ref{fig_TwoPatch_185_160}) and a second minimum at $\Psi \approx
1$.  This second minimum corresponds to a crystal state with all
the patches involved in contacts. Thus we see that this
double-well free energy describes a liquid-solid phase transition
rather than to a monomer-(one dimensional) fiber transition.  This
liquid-solid transition is what one would expect for this model.

\begin{figure}
\center
\includegraphics[width=8cm]{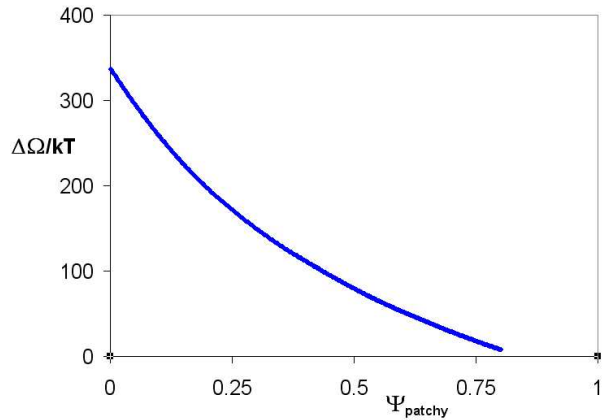}
 \caption{Plot of the free energy $\Delta \Omega$ in units of $kT$ versus the order
 parameter for the patches, $\Psi$,  defined in the text.  Simulations are
 at a temperature $kT=.185$ and pressure $P=2.35$.  The minimum at $\Psi$ in the vicinity
  of $0.9$ is the crystal state.}
 \label{fig_TwoPatch_185_235}
\end{figure}

\noindent Thus, as expected, there is no  nucleation mechanism for
the monomer-fiber transformation. In the case of $T=0.185$ and
$P=2.35$ the system simultaneously crystallizes and increases its
number of patch contacts. The system successfully reaches an
equilibrium crystal state and the free energy has only one minimum
at this state, as seen in  Figure \ref{fig_TwoPatch_185_235}.
However, at lower temperature the picture is quite different. At
$T=0.1$ and $P=0.01$ the fibers form \textbf{before}
crystallization can occur. The free energy has one minimum at
$\Psi$ around $0.9$, but the system is not crystallized. As the
set of fibers is formed, the dynamics slows down significantly and
the system becomes stuck in a non-equilibrium state. Figure
\ref{fig_HbS_Chains1} shows an example of a typical configuration
for this nonequilibrium state, corresponding to a set of rods in a
"glassy" state.

The umbrella sampling simulations were performed on a system of N
particles, with $N$ = 500, in the NPT ensemble, with a range of
interaction given by $\lambda=1.25.$  The equation of state
simulations were also performed in the NPT ensemble with 500
particles. The details of the umbrella sampling technique can be
found in \cite{Frenkel_92_01,Frenkel_96_01}. In short, this method
is based on a biasing of the original interactive potential in
such a way that the system is forced to reach otherwise
inaccessible regions of the configuration space. In particular,
for simulations starting in a liquid state, the system is forced
towards larger values of the order parameter. Since the actual
dependence of the free energy on the order parameter (for the
entire range of values of $\Psi$) is not known, the biasing
function is chosen to be quadratic:
\begin{equation}
U_{biased}(\textbf{r}^N) = U_{unbiased}(\textbf{r}^N) +
k(\Psi(\textbf{r}^N) - \Psi_0)^2 \notag
\end{equation}
where the parameters $k$ and $\Psi_0$ determine which region of
values of $\Psi$ would be sampled in the simulation. By changing
these parameters we can sample the entire region of values of
$\Psi$. An interesting result of the simulations is that at
intermediate pressures the system when biased to large values of
$\Psi$ starting from an initial liquid state has a very large
volume. However, if the system is started from an initial (fcc)
crystal state, the volume of the system remains small (still in a
crystal state), while the order parameter value is around 1. This
observation suggests that at not very high supersaturation the
biased system started from the fluid  ends up by forming a few
long fibers, rather than a set of fibers that are packed into a
crystal lattice. The fiber formation time therefore is much
smaller than the crystallization time at low supersaturation.
Starting from aa fcc initial condition, however, the particles
just reorient within the crystal lattice to form the fibers,
remaining in the crystal state. At higher pressure, and therefore
higher supersaturation, (such as in Figure
\ref{fig_TwoPatch_185_160}), the crystallization occurs in a time
comparable with the fiber formation time. While the system is
forced to form the fibers, these fibers pack into the crystal
lattice.

\subsection{Equation of State}

\begin{figure}
\center
\rotatebox{-90}{\includegraphics[width=8cm]{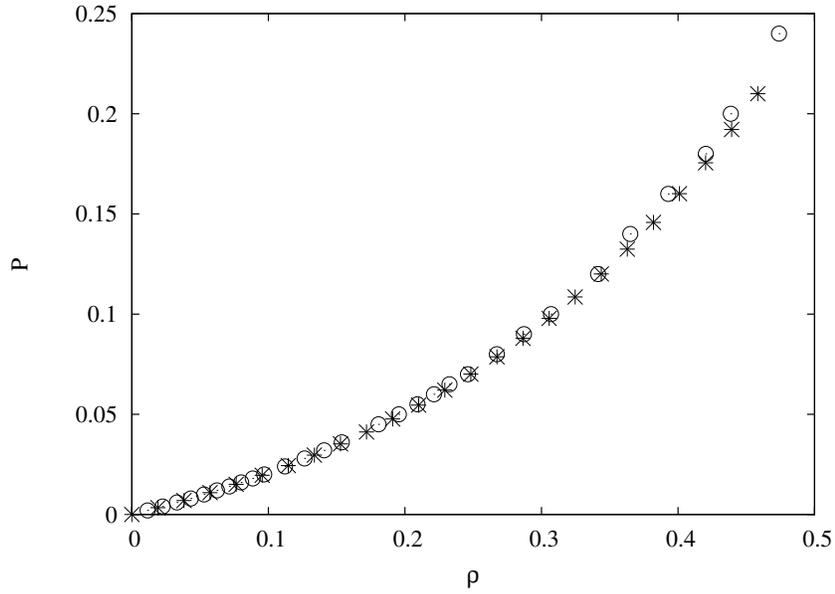}}
 \caption{A plot of the pressure,p,versus density, $\rho$, in the fluid phase at $kT=.17$( in units
 of the well depth $\epsilon$).   The open
 circles are the results of the simulation.  The asterisks are theoretical results obtained
 from Sear \protect\cite{Sear_99_01}. The number of patches $m=2$, with a
 patch angle of about $\delta=52$ degrees and a range of
 interaction$\lambda=1.25$, i.e.
 $r_c=1.25 \sigma$, where $\sigma$ is the hard core diameter.}
 \label{fig_EqState_kT.17}
\end{figure}

\begin{figure}
\center
\rotatebox{-90}{\includegraphics[width=8cm]{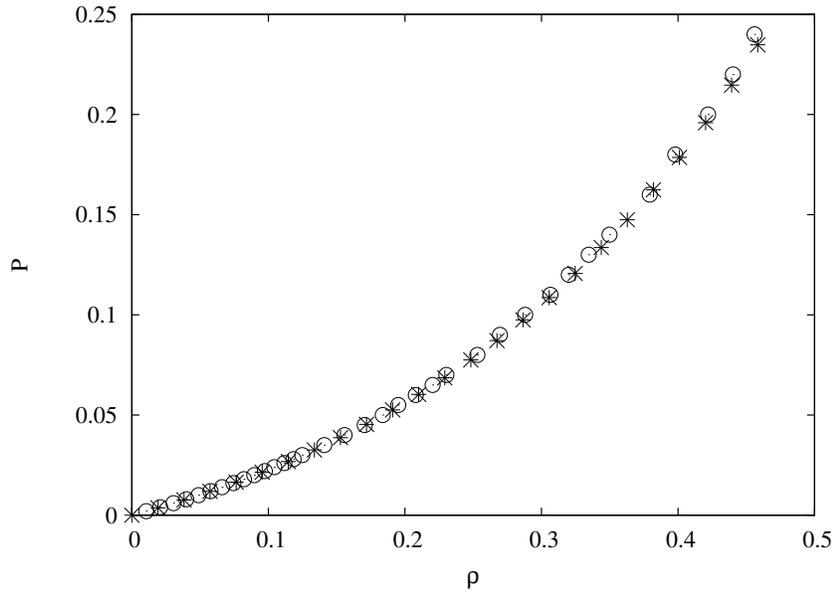}}
 \caption{A plot of the pressure versus density in the fluid phase at $kT=.185$  The open
 circles are the results of the simulation.  The asterisks are theoretical results obtained from
 Sear \protect\cite{Sear_99_01}.  Same values for parameters as in Figure \ref{fig_EqState_kT.17}.}
 \label{fig_EqState_kT.185}
\end{figure}

The equation of state for the two patch model is shown for the two
low temperatures studied in Figures \ref{fig_EqState_kT.17} and
\ref{fig_EqState_kT.185}.

\begin{figure}
\center
\rotatebox{-90}{\includegraphics[width=8cm]{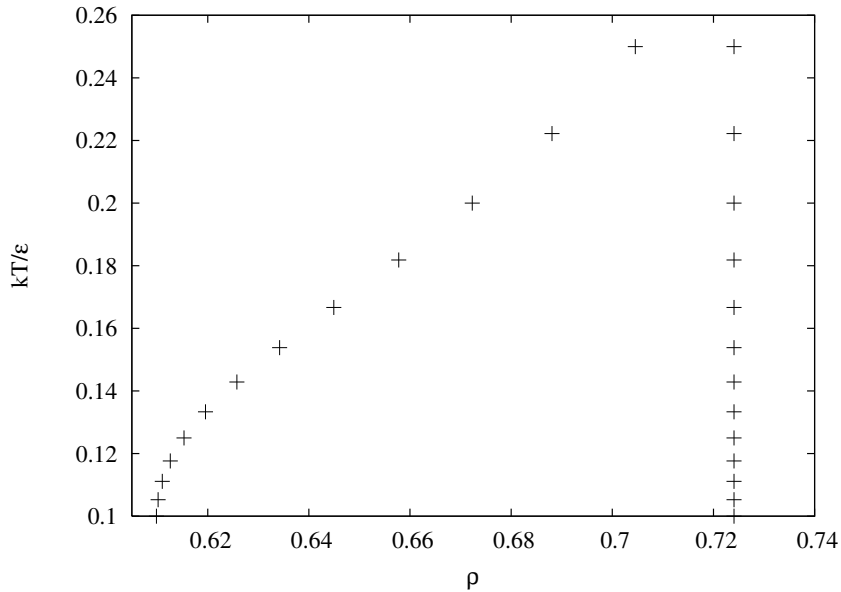}}
 \caption{Theoretical prediction for the phase diagram for the two
 patch model discussed
 in the text.  The theory is due to Sear
\protect\cite{Sear_99_01}. Same values for parameters as in Figure
\ref{fig_EqState_kT.17}.
 }
 \label{fig_PhaseDiagram_n=2}
\end{figure}

\noindent Also shown in these figures are results  from a  theory
for the m site model \cite{Jackson_88_01} of globular proteins due
to Sear \cite{Sear_99_01}. This model is identical to our patch
model defined in section 2 for the case in which the various
interaction parameters are equal.  Sear studied the m-site model
(m patches) using the Wertheim perturbation theory
\cite{Wertheim_87_01} for the fluid phase and a cell model for the
solid phase \cite{Vega_98_01} and showed that the model exhibits a
fluid-fluid transition (for $m > 2$) which is metastable with
respect to the fluid-solid transition for most values of the model
parameters. For $m=2$, however, there is only a fluid-solid
transition. As can be seen from Figures \ref{fig_EqState_kT.17}
and \ref{fig_EqState_kT.185}, the theory yields results which are
in excellent agreement with the results of our simulation. For
completeness we show the theoretical prediction for the
fluid-solid transition for $m=2$  in Figure
\ref{fig_PhaseDiagram_n=2}, assuming a fcc crystal structure.

\begin{figure}
\center
\rotatebox{-90}{\includegraphics[width=8cm]{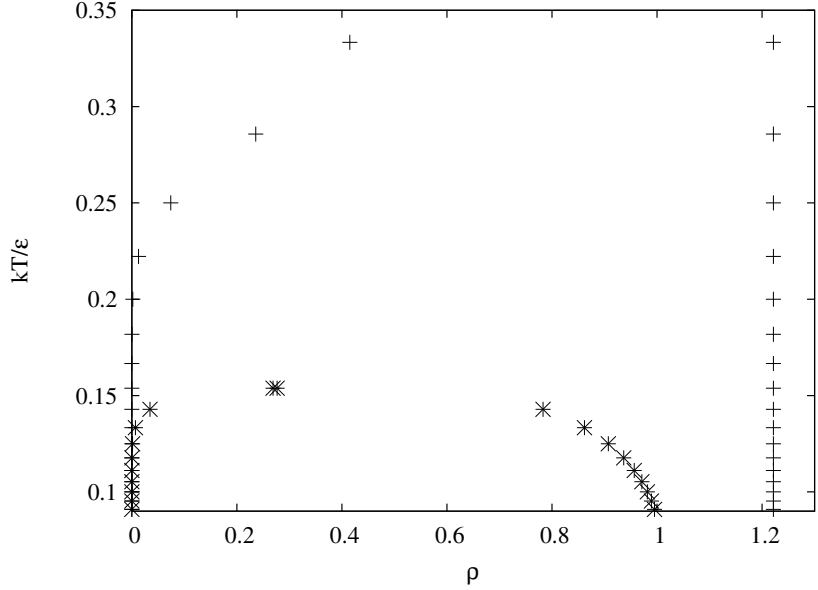}}
 \caption{Theoretical prediction for phase diagram for the model of HbS discussed
 in the text, in which all interaction parameters are equal. Theory is due to Sear
\protect\cite{Sear_99_01}. Here $m=8$, $\lambda=1.05$ and
$\delta=51$ degrees. The fluid-fluid transition
 is metastable.}
 \label{fig_PhaseDiagram_n=8}
\end{figure}

We also show in Figure \ref{fig_PhaseDiagram_n=8} the results of
the theory for the case $m=8$, as the model discussed in section 2
has 8 pairs of interacting patches.  Since the interactions
between the different sites in the model studied by Sear are
assumed to be equal, the model lacks the anisotropy discussed in
section 2 that is necessary to account for the  fiber formation in
HbS molecules. Nevertheless, it is quite instructive to know the
phase diagram for this case.

\noindent The fluid-fluid binodal curve has an upper critical
point for $m>2$ (e.g. Figure \ref{fig_PhaseDiagram_n=8}),unlike
the case for HbS. In that case experimental measurements by Palma
et al \cite{Palma_91_01} display a spinodal curve.  Such a
spinodal implies the existence of a binodal curve with a lower
critical point . However, as shown by Shiryayev et al
\cite{Shiryayev_05_01} the lower critical point reflects the
crucial role of the solvent in the case of HbS in solution.  The
solvent is not taken into account in our model, but presumably if
one would include a solute-solvent coupling similar to that of
\cite{Shiryayev_05_01}, this coupling could change the phase
diagram shown in Figure \ref{fig_PhaseDiagram_n=8} to one with a
lower critical point, as found by Shiryayev et al
\cite{Shiryayev_05_01}, e.g. Figure \ref{fig_solvent}.

Finally, we note that  Jackson et al \cite{Jackson_88_01} used
Wertheim's theory for the two site model to predict that the
fraction of molecules that are present in chains of length n is
given by
\begin{equation}
nX^{2}(1-X)^{n-1}
\end{equation}
while the average change length, <n>,  is given by
\begin{equation}
<n> = 1/X.
\end{equation}
 Here $X$ is the fraction of sites that are not bonded to
another site and is given by \cite{Sear_99_01}
\begin{equation}
X=\frac{2}{1+[1+4\rho Kg_{hs}^{c}\exp(\beta\epsilon)]^{1/2}}
\end{equation}
where $g_{hs}^{c}$ is the contact value of the pair distribution
function for a fluid of hard spheres. The quantity $K$ is given by
the expression
\begin{equation}
K =\pi\sigma^{2}(r_c-\sigma)(1-\cos(2\delta))^{2}.
\end{equation}

\begin{figure}
\center
\rotatebox{-90}{\includegraphics[width=8cm]{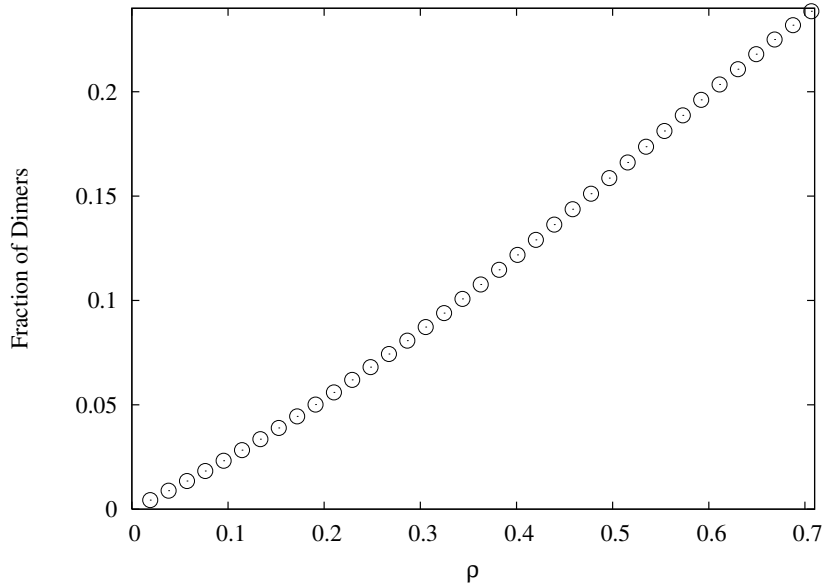}}
 \caption{Theoretical prediction for the fraction of dimers as a function of density
 at $kT=.185$. From \protect\cite{Jackson_88_01}.}
 \label{fig fraction_dimers}
\end{figure}

\begin{figure}
\center
\rotatebox{-90}{\includegraphics[width=8cm]{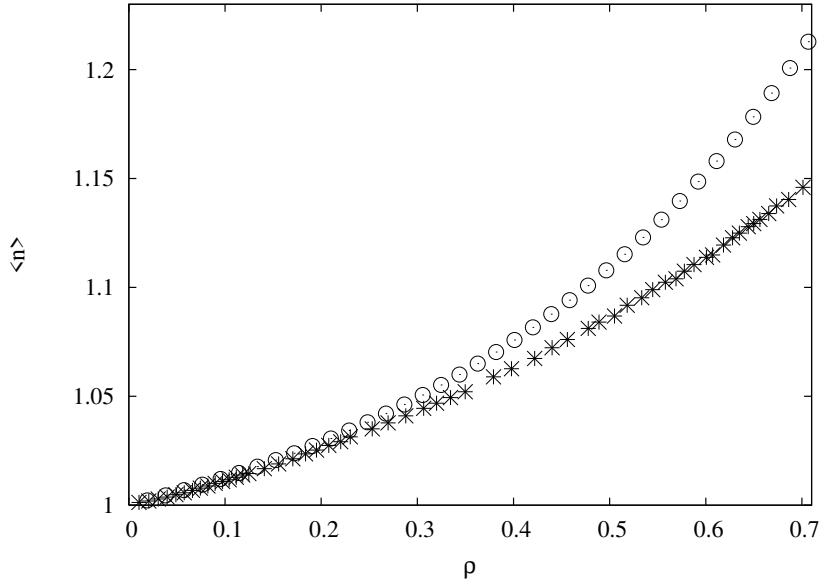}}
 \caption{Theoretical prediction for the average chain length as a function of
 density at $kT=.185$ (open circles). Also shown are the simulation results for
 $1/(1-\Psi$)
 (asterisks)-see discussion in text. }
 \label{fig_chain_length}
\end{figure}

\noindent For example, the theoretical prediction for the fraction
of dimers and the average chain length as a function of density
(at $kT=.185$) is shown in Figures \ref{fig fraction_dimers} and
\ref{fig_chain_length}, respectively. Also shown for comparison in
Figure \ref{fig_chain_length} is an approximation for the average
chain length obtained from our simulation results for the order
parameter.  Had we chosen to use a function $f=1$ in our
definition in section 3 (rather than $f(x)=x$ for $x>\cos\delta$)
the order parameter would have been equal to $1-X$, i.e. the
fraction of sites that are bonded to another site. In that case
the average chain length would be equal to $1/(1-\Psi)$. As shown
in the figure, $1/(1-\Psi)$ is in general less than the average
chain length, due to  our choice of f in section 2.


\section{Conclusion}
The results obtained in the previous section raise several
important questions. For some thermodynamic conditions the system
crystallizes and the fibers align along each other to form a fcc
like structure.  For other thermodynamic conditions the fiber
formation prevents the system from crystallizing and it remains in
a non-equilibrium glassy state. Is there some boundary between
these two behaviors? In protein crystallization science a somewhat
similar phenomenon is known as  "gelation". The boundary between
successful crystallization and the nonequilibrium "gel" state is
the "gelation line". (This line is obviously not an equilibrium
phase boundary.) This "gel" state is not a gel in the usual sense.
It is more like a glassy state in which the aggregates form,  but
due to their formation the dynamics slows down significantly and
the aggregates cannot subsequently form crystals.  Thus the system
becomes stuck in this glassy state. Recently mode coupling theory
has been used to predict the gel line for protein crystallization
\cite{Kulkarni_03_01}. It is possible that one can use the same
approach for the two patch model.

Another interesting question arises when we compare the behavior
of the two-patch model and  real sickle hemoglobin . Ferrone's
studies \cite{Ferrone_02_03,Ferrone_85_01} show that the formation
of the fourteen strand fibers occurs via a nucleation mechanism.
Molecules aggregate into an initially isotropic droplet which
subsequently becomes an anisotropic fiber. The kinetics of the
transformation of the isotropic  droplet into an anisotropic fiber
is not well understood. It is believed that this happens through
the attachment of molecules to  active sites of the molecules in
the fiber. At some fiber diameter  the number of active sites on
the fiber surface  is not sufficient to induce layer by layer
growth of the fiber in the direction perpendicular to the fiber
axis. Thus molecules that attach to the active sites form a
droplet that then detaches from the fiber, with the subsequent
formation of a new fiber (i.e. the new fiber forms via
heterogeneous nucleation from the original fiber). This process
explains why an  anisotropic fiber does not continue to increase
its diameter.

The point of this discussion is to note that one of the main
differences between the two-patch model and sickle hemogloblin
molecules is that the latter forms fibers through nucleation,
whereas the former does not. One way to improve the two patch
model is to increase the number of patches. If one includes an
additional two  active patches, corresponding to the lateral
contacts, then most likely this system would form non-interacting
double strands.  If so, this would not lead to any qualitative
difference with the two patch model. Another approach would be to
add several relatively weak patches around the particle which can
represent inter strand interactions. However, this would not
necessarily give  two distinct  nucleation mechanisms
(monomer-fiber and fiber-crystal). Rather, it is more likely that
the crystallization would be anisotropic. In order to produce a
nucleation from monomers to fibers it might be necessary to have a
particular distribution of patches such that at some radius of the
anisotropic droplet (pre-fiber) its growth in the radial direction
is  significantly depressed.

\section{Acknowledgements} This material is based upon work
supported by the G. Harold Mathers and Leila Y. Mathers Foundation
and by the National Science Foundation,  Grant DMR-0302598.

\end{document}